\documentstyle[epsfig]{mn}
\begin{document}
\def\ltsima{$\; \buildrel < \over \sim \;$}
\def\simlt{\lower.5ex\hbox{\ltsima}}
\def\gtsima{$\; \buildrel > \over \sim \;$}
\def\simgt{\lower.5ex\hbox{\gtsima}}

\title[On the nature of the X-ray absorption in Seyfert~2 galaxies]
{On the nature of the X-ray absorption in Seyfert~2 galaxies}

\author[M. Guainazzi, et al.]
{M. Guainazzi$^{1,2}$, F.Fiore$^{3,4}$, G.Matt$^5$, G.C.Perola$^5$\\ ~ \\
$^1$Astrophysics Division, Space Science Department of
ESA, ESTEC, Postbus 299, NL-2200 AG Noordwijk, The Netherlands \\
$^2$XMM Science Operation Center, VILSPA-ESA, Apartado 50727, E-28080 Madrid, Spain \\
$^3$Osservatorio Astronomico di Roma, Via dell'Osservatorio, I-00144 
Monteporzio Catone, Italy \\
$^4$BeppoSAX Science Data Center, Via Corcolle 19, I-00131 Roma, Italy \\
$^5$Dipartimento di Fisica, Universit\`a degli Studi Roma 
Tre, Via della Vasca Navale 84, I-00146 Roma, Italy \\
}

\maketitle
\begin{abstract}

We have studied the correlation among X-ray absorption, optical
reddening and nuclear dust morphology in Seyfert 2 galaxies.
Two main conclusions
emerge: a) the Balmer decrement and the amount of X-ray absorption are
anticorrelated on a wide range of column density: ${\rm
10^{21} \simlt N_H \simlt 10^{24}}$~cm$^{-2}$. The correlation does no
longer apply to
Compton-thick objects (${\rm N_H \simgt 10^{24}}$~cm$^{-2}$), although
they span a comparable range in Balmer decrement;
b) Compton-thin Seyfert 2s seem to prefer nuclear environments,
which are
rich of dust on scales of the hundreds parsecs.
On the other hand, Compton-thick
Seyferts exhibit indifferently
``dust-poor'' and ``dust-rich'' environments. These results support
an extension of the Seyfert unification scenario (as recently
proposed by Matt, 2000), where Compton-thick Seyfert 2s are observed
through compact ``torii'', whereas Compton-thin ones are
obscured by dust on much larger scales

\end{abstract}


\section{Introduction}

The nuclear X-ray emission of most Seyfert~2 galaxies is seen
through substantial
amount of neutral absorbing matter (Awaki et al. 1991; Turner et al. 1997). This
is in good agreement with the predictions of the ``Seyfert unified
theories'' (Antonucci \& Miller 1985; Antonucci 1993), which were originally
invoked to explain the broad optical lines emerging in the
spectropolarimetric observations of NGC~1068 (Antonucci
\& Miller 1985) and of several other Seyfert~2s (Tran 1995; Heisler et al.
1997). These theories postulate that Seyfert~2 nuclei are seen through a dusty
molecular torus, which surrounds the nuclear environment on scales of the
order of about 1~pc. Even those few cases, where no X-ray
absorption is measured,
can be easily reconciled with
this scenario. The very flat X-ray spectra and huge fluorescent iron lines
suggest that the nucleus is completely hidden in the 2--10~keV
and that the nuclear radiation
is seen via reflection or electron scattering
(Matt et al. 1996a). Well known examples of this
phenomenology are Circinus Galaxy (Matt et al. 1996b), NGC~7674 (Malaguti
et al. 1997) and NGC~6240 (Iwasawa \& Comastri 1998), along with NGC~1068
itself (Iwasawa et al. 1997; Matt et al. 1996b).
Recent BeppoSAX (Boella et al. 1997) observations
have unveiled strongly absorbed
($N_H \simgt$ a few~$10^{24}$~cm$^{-2}$) components
above 10~keV in the Circinus Galaxy (Matt et al. 1999),
and NGC~6240 (Vignati et al. 1999). It is straightforward
to interpret these components as the direct view of
the nuclear emission, which is mirrored in the 2--10~keV
band.

In this paper, following the most common nomenclature,
we will refer to the above classes of Seyfert~2s as {\it Compton-thin}
and {\it Compton-thick}, respectively.

According to the unification scenario, the difference between
Compton-thin and -thick objects is mainly in
the value of the column density, ${\rm N_H}$.
If it exceeds $\simeq$10$^{24}$~cm$^{-2}$, the
matter becomes thick to Compton scattering, the impinging photons are
down scattered to energies where the photoabsorption cross section is
dominant, the nuclear radiation is totally suppressed and we observe a
Compton-thick object. If the column density has lower values (either
because the matter is less dense or due to a shorter optical path to
the observer), part of the incoming radiation can be transmitted through
the absorbing cloud(s).

Two recent papers provide for the first time a conspicuous basis to study the
X-ray properties of a sizable sample of Seyfert~2 galaxies and their
correlation with their multiwavelenght spectral energy distribution and
nuclear morphology. Bassani et al. (1999, B99) have recently
collected all the Seyfert~1.8 to 2s,
for which reliable hard X-ray spectra are
available, and listed the corresponding absorbing column
densities, together with [O{\sc iii}] luminosities and
Balmer decrements.
This catalogue makes use of all the available X-ray
spectroscopic measurements up to ASCA and BeppoSAX
to discriminate between nuclear transmitted
and scattered or diffuse emission.
It should therefore allow a statistically robust
identification of the proper absorbing column density
towards the nucleus.
By construction, this sample is not unbiased, the main
potential selection effect being against obscured objects at the
faintest fluxes. Risaliti et al. (1999), however, suggest that such a bias
can be strongly reduced if only sources with an [O{\sc iii}] flux
higher than $3 \times 10^{-13}$~erg~cm$^{-2}$~s$^{-1}$ are
considered. Underlying this statement, it is assumed
that the [{O{\sc iii}] luminosity is a good ({\it i.e.}:
within a factor of a few) estimator of
the intrinsic nuclear power (Maiolino et al. 1998).

On the other hand,
Malkan et al. (1998, M98) carried out a systematic slew survey of
a sizable sample of nearby galaxies with the WFPC2 on board the
{\it Hubble Space Telescope} (HST), aiming at studying
the nuclear morphology in Seyfert~1, Seyfert~2 and H{\sc ii}
galaxies. Their images unveiled a plethora of complex structures on scales
of the hundreds of parsecs, both in emission and in absorption. A summary of
their findings and classification 
is given in Sect.~2.

\subsection{The scope of the paper and its current observational context}

In this {\it paper} we aim at studying the correlation between X-ray
absorption, optical reddening and nuclear dust morphology.
Our results suggest that
the absorbing matter in Compton-thick and -thin Seyfert~2s
is associated with different physical regions or geometries, and
propose that a correlation with the absorption structures
seen by the HST/WFPC2 may exist in Compton-thin Seyferts only.
The interpretation of the results must take into account
the limitations of the X-ray instruments, whose measurements have
been employed to build up the B99 catalogue. At the level of spatial
resolution ($\sim$~arcminutes) of such detectors in the
2--10~keV energy band,
one needs to {\it assume} that the bulk of the observed
X-rays comes from a region very close
to a nuclear engine powered by accretion onto a
supermassive black hole.
The column density derived from the spectral fits
provides an integrated measurement of the
amount of absorbing matter along the line of sight,
which we associate with a geometrically
thin, possibly not homogeneous
screen. If the X-ray
emission is due to the superposition of different
components, emitted by gas at different scales/distances
from the active nucleus, the integrated
spectrum could be misleading. Moreover, each of these components could be
differently absorbed. Even if the absorption structure
along the line of sight is known, little can in
principle be said on the extent and distribution of the gas
along different viewing angles, and mostly from
variability arguments (see, {\it e.g.} Guainazzi et al. 2000).

However, a nuclear origin for the bulk of the
X-ray emission above 2~keV
in Compton-thin Seyfert~2 galaxies can
hardly be challenged. Their X-ray luminosity
(Smith \& Done 1996; Turner et al. 1997),
variability properties (see, {\it e.g.}, Guainazzi et al. 1998),
and the very presence
of nuclear point-like sources in the ROSAT images
(alongside with extended emission in a few objects;
cf.
Matt et al. 1994; Morse et al. 1995; Weaver et al. 1995;
Rush \& Malkan 1996) strongly support
the existence of a dominating nuclear source, seen either
in transmission, through scattering, or both.
The discovery of transmitted, highly
absorbed components above 10~keV
in a few scattering-dominated Seyfert~2s provides a
strong observational support for the
application of the same idea to
Compton-thick objects as well.
The first results provided by the {\it Chandra} observatory
(whose payload allows X-ray imaging
with an unprecedented
spatial resolution of $\simeq$0.5") broadly support
the same scenario. Most of the emission in Mkn~3 (Sako et al.
2000) and the Circinus~Galaxy (Sambruna et al. 2001) is concentrated
in an unresolved source, coincident with the active nucleus.
In the Circinus~Galaxy
the upper limit on the size of the unresolved emitting region is
$\sim$15~pc only. Its spectral properties
are very close to those observed by ASCA
(Matt et al. 1996) or BeppoSAX (Matt et al. 1999; Guainazzi et al. 1999).
This unresolved component is often embedded
in diffuse gas, whose contribution to the total emission
substantially decreases
with energy. Conversely, only diffuse emission (on scales $\sim$165~pc)
is observed in NGC~1068 (Young et al. 2001), where the
column density to the nucleus is likely to exceed $10^{25}$~cm$^{-2}$
(Matt et al. 1997), and therefore no transmission through the
absorber is expected. Although these data are opening
new exciting perspectives to the study of the distribution
of matter in the nuclear region of nearby Seyfert~2s, they do
not contradict the scenario depicted before the advent
of {\it Chandra}, and which is assumed throughout this
{\it paper}

The paper is organized as follows.
Sect.~2 briefly summarizes the M98 sample properties.
In Sect.~3, we present the ``Balmer Decrement versus X-ray absorption'' plane.
The correlation of the M98 nuclear morphology classes with several
others physical and morphological observables is discussed in Sect.~4.
In Sect.~5 and 6 we briefly discuss some implications of our findings.
The main results of this {\it paper} are summarized in Sect.~7.

\section{The nuclear morphology sample}

The sample of M98 contains 256 of the nearest
(${\rm z < 0.04}$, see Fig.~\ref{fig1}) Seyfert~1s (91), Seyfert~2s (114)
\begin{figure}
\begin{center}
\epsfig{figure=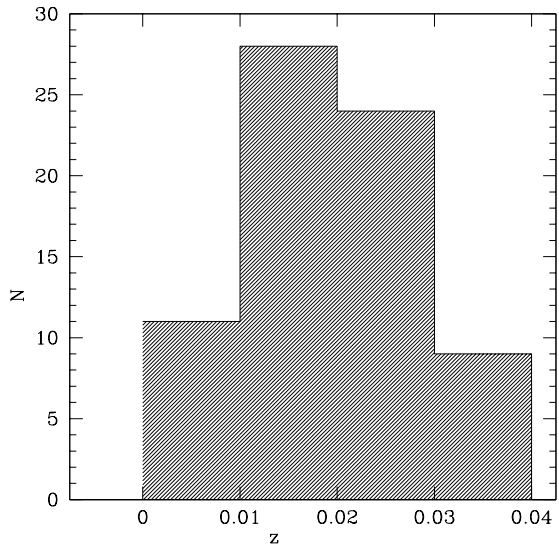,height=8.0cm,width=8.0cm}
\end{center}
\caption{Redshift distribution of the Malkan et al. (1998) sample. Only
objects for which measures of both O{\sc [iii]} and FIR luminosity are
available, allowing direct comparison with Figures~3 and
4}
\label{fig1}
\end{figure}
and starburst galaxies (51). For each galaxy images with the F606W filter
on the WFC2 on board the HST were taken,
which has a mean wavelength of 5940${\rm \AA}$ and
a Full Width Half Maximum of 1500${\rm \AA}$. The plate scale of 0''.046 per
pixel corresponds to 130~pc for the farthest object. A wealth of different
structures are seen in these images, both in emission (bars, rings,
filaments/wisps) and in absorption (dust lanes and patches).
Without entering in the
details of the classification (which is admittedly qualitative and
partly subjective) the
most relevant result to our paper
is that Seyfert~2 galaxies are significantly more likely to
exhibit dusty nuclear
environments (see, however, Antonucci 1999 for a different view).
M98 speculate that
``this galactic dust could produce much of the absorption in Seyfert~2
nuclei which had instead been attributed to a thick dusty accretion torus
forming the outer part of the central engine'' and elaborate an alternative
model to account for the observed absorption in term of galactic dust
(Galactic Dust Model, GDM).
Given the size of the M98
sample and the need for homogeneous experimental conditions
to assess the morphological
properties in the most unbiased way,
we have not tried to include other smaller imaging surveys in our study,
as that of Carollo et al. (1997).

\section{The X-ray absorption versus Balmer Decrement plane}

In Fig.~\ref{fig2} we plot the Balmer Decrement (BD) versus the
\begin{figure}
\begin{center}
\epsfig{figure=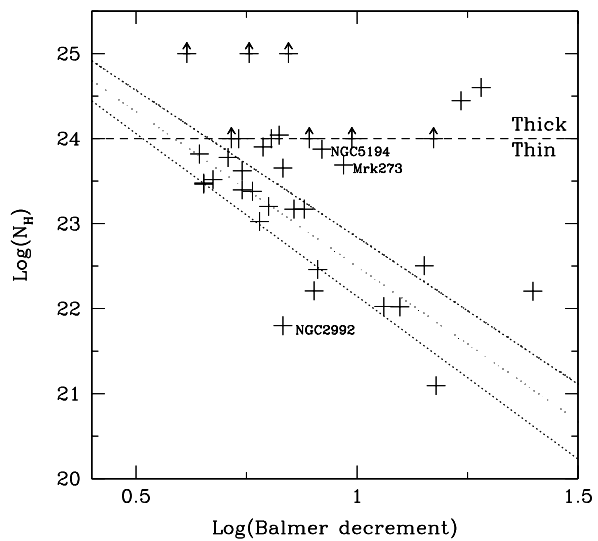,height=8.0cm,width=8.0cm}
\end{center}
\caption{Balmer decrement (BD) versus X-ray absorption column density
plane for Seyfert~2 galaxies.
The {\it dotted lines}
mark the 95\% confidence interval for the correlation between the
two quantities, which holds for objects with ${\rm N_H}$ comprised
between $10^{21}$ and $10^{24}$~cm$^{-2}$. The statistical
uncertainties on the ${\rm N_H}$ measures are not plotted for
clarity, but used to calculate all the correlation and
statistical quantities derived from the plot. The {\it
dashed line} separate Compton-thin and -thick objects}
\label{fig2}
\end{figure}
X-ray absorption column density for a sub-sample of Seyfert~2s extracted
from the B99 catalog.
The sub-sample includes only:
a) objects whose redshift is $<$0.04, to be consistent
with the M98 sample; b) objects with ${\rm f_{[OIII]} > 4 \times
10^{-13}}$~erg~cm$^{-2}$~s$^{-1}$,
as suggested by Risaliti et al. 
(1999) to ensure the completeness of the B99 sub-sample.

The plot is rather complex and exhibits some interesting features. Let's
consider first the Compton-thin sources. A strong {\it anticorrelation} between
Balmer decrement 
and ${\rm N_H}$ is evident. The linear correlation coefficient is
-0.64 for 19 objects, corresponding to a chance occurrence likelihood
$\simlt$0.5\%. The dotted lines
in Fig.~\ref{fig2} represent the loci of the plain delimited by
the liner best-fit to the Compton-thin data points,
when the 95\% statistical uncertainties
on the best-fit parameters are taken into account. The widest deviations
from the correlation are represented by two objects with high Balmer
decrement and
high ${\rm N_H}$ (Mkn~273 and NGC~5194), and
NGC~2992, which exhibits an X-ray absorption about one order of magnitude
lower than expected on the basis of its Balmer decrement.
In the former case, the available measures of X-ray absorption are
poorly constrained and consistent with the correlation.
On the other hand, NGC~2992 exhibits secular variations of the X-ray flux
by a factor up to 20 (Weaver et al. 1996; Gilli et al. 2000).
It may well be that such a peculiar behavior affects
the long-term properties of the nuclear
emission of this objects in other wavelengths. 

This anticorrelation, whatever is its origin, is {\it not} valid for the
Compton-thick objects. They span the same range of Balmer
decrement values as the
Compton-thin ones, while, according to the above anticorrelation, they
should lie preferentially  at much lower values than the lowest
abscissa value
in Fig.~\ref{fig2}. For instance, the Circinus~Galaxy and NGC~6240
(among the first Compton-thick
Seyfert~2s, whose
column density has been reliability measured; Matt et al.
1999; Vignati et al. 1999), have a value of
${\rm N_H}$ three orders of magnitude higher than expected from the
Balmer decrement
versus ${\rm N_H}$ correlation valid for the Compton-thin objects.

The same plot as in Fig.~\ref{fig2} is shown in Fig.~\ref{fig7} for
\begin{figure}
\begin{center}
\epsfig{figure=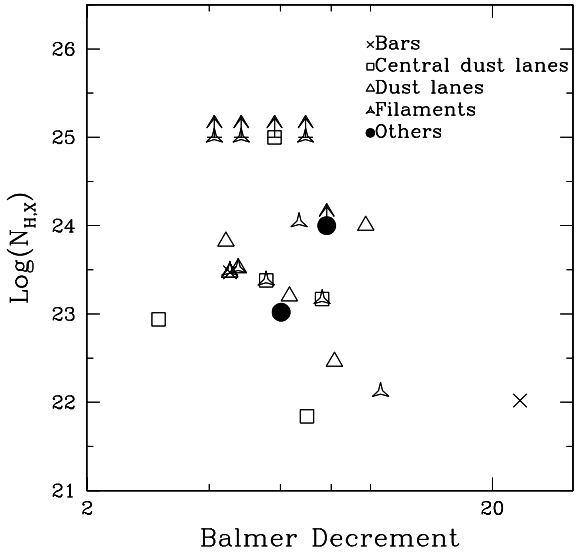,height=8.0cm,width=8.0cm}
\end{center}
\caption{The same plot of Fig.~2, when only the sources for which a
nuclear morphology classification after M98 exists are considered.
Different symbols indicate the M98 main classification classes.}
\label{fig7}
\end{figure}
the Seyfert~2s only, which have been observed by M98. The M98 classification
is reported (eventually replicated with superimposed markers if more
than one structure is present simultaneously)\footnote{We remind here that
the nuclear features classified by
M98 are not mutually exclusive. A galaxy can exhibit simultaneously, for
instance, dust
lanes and emitting filaments}.
An eye inspection of Fig.~\ref{fig7} seems to suggest that galaxies
\begin{figure}[h]
\begin{center}
\epsfig{figure=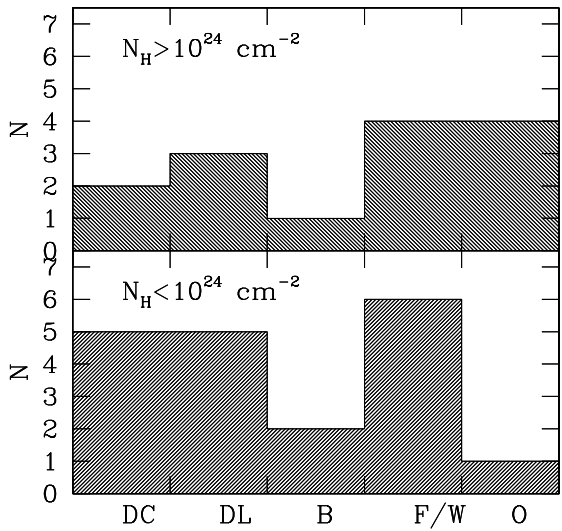,height=8.0cm,width=8.0cm}
\end{center}
\caption{Distribution histogram for the morphology classes in M98
for Compton-thick (${\rm N_H > 10^{24}}$~cm$^{-2}$, {\it
upper panel}) and Compton-thin (${\rm N_H \le 10^{24}}$~cm$^{-2}$, {\it
lower panel}) objects. DC~=~central dust lanes; DL~=~non-central dust lanes;
B~=~bars; F/W~=~filaments/wisps; O~=~others}
\label{fig3}
\end{figure}
whose nuclear environment is dust-rich are more likely to be found in
the Compton-thin half-plane, while dust-poor ones are more common in the
Compton-thick half-plane. This hypothesis is investigated quantitatively
in Fig.~\ref{fig3}, where the distribution function of the M98 morphology
classes is reported for Compton-thin and -thick objects
separately. The ratio of dust-rich versus dust-poor
objects is 5:6 ($0.8 \pm 0.4$) and 10:3 ($3.3 \pm 2.1$) for Compton-thick
and Compton-thin objects, respectively.
This suggests that Compton-thick Seyfert~2s
can be equally found in ``dust-rich'' or ``dust-free'' environments,
while Compton-thin ones exhibit a marked preference for ``dust-rich''
environments. Although the nominal value of this ratio is three times
higher in Compton-thin than in Compton-thick objects, the sample is
still too small to allow us to draw firm conclusions on this points.
An enlargement of the sample, for which high-resolution imaging and reliable
X-ray spectra are available,
is urgently needed, and can be possibly achieved in the
next future thanks to XMM-Newton.

\section{Correlation with the nuclear dust morphology classes}

Although the evidence for a systematic difference in the nuclear dust
content between Compton-thick and -thin Seyfert~2s is still marginal, we
have tried to correlate the M98 morphology classes with several intrinsic
nuclear power and/or reprocessing indicators.
Some results of this study are shown in Fig.~\ref{fig5}.
\begin{figure*}
\begin{center}
\epsfig{figure=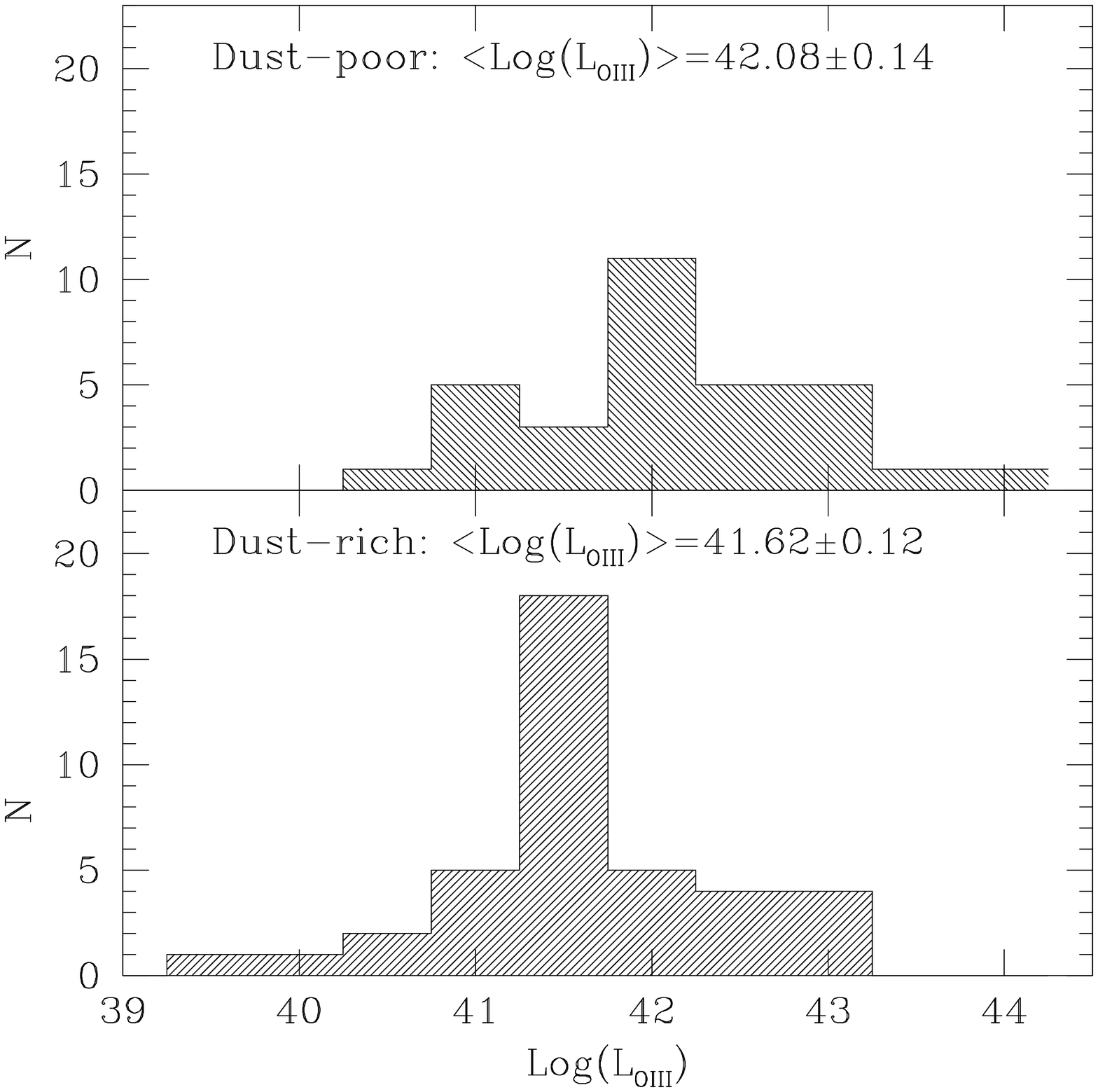,height=8.0cm,width=8.0cm}
\epsfig{figure=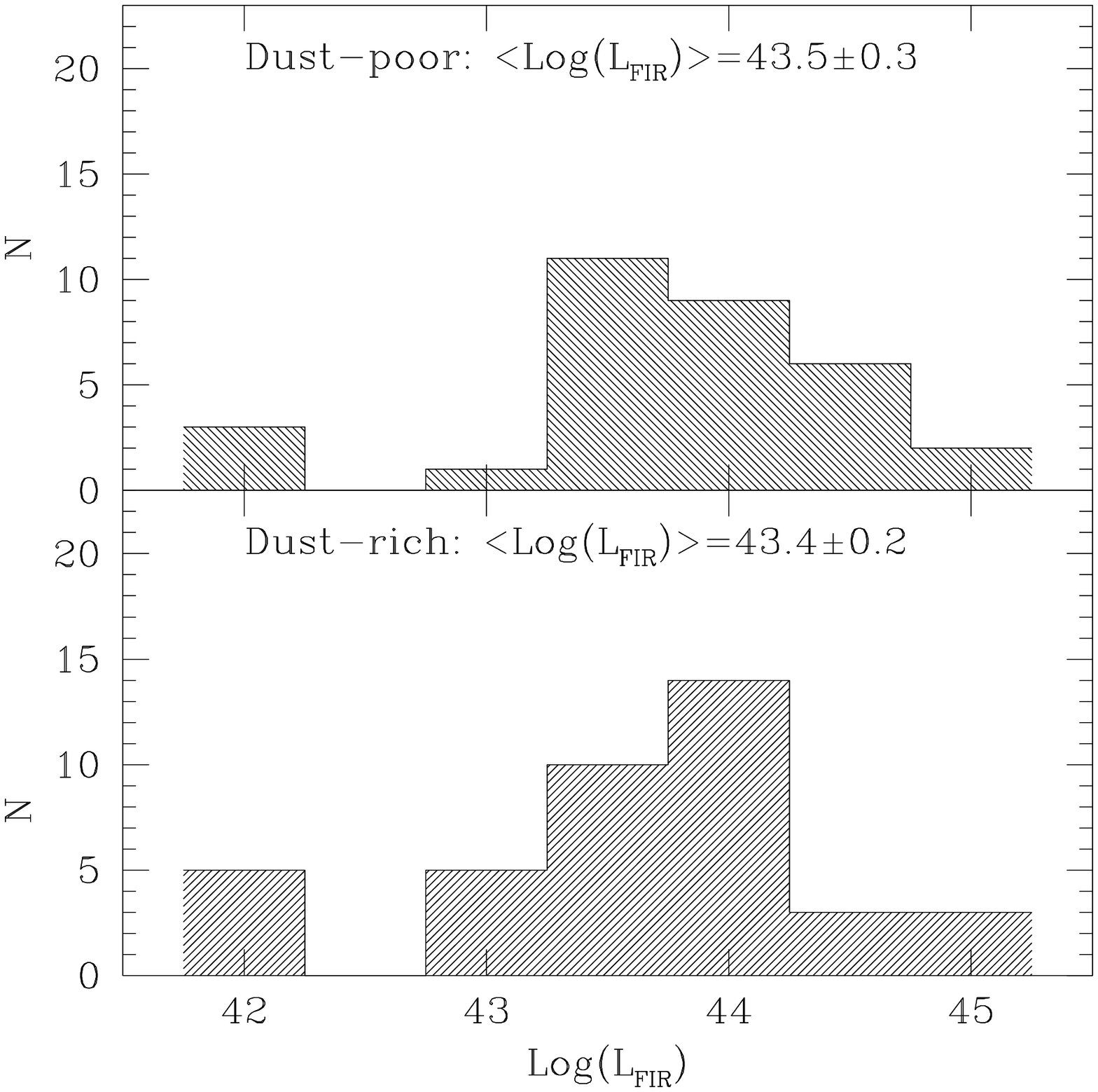,height=8.0cm,width=8.0cm}
\end{center}
\caption{
{\it Left upper}:
distribution histogram of [O{\sc iii}] luminosities
for ``dust-poor'' ({\it upper panel}) and ``dust-rich'' ({\it lower panel})
objects;
{\it Right upper}: distribution histogram of FIR luminosities
for ``dust-poor'' ({\it upper panel}) and ``dust-rich'' ({\it lower panel})
objects}
\label{fig5}
\end{figure*}
The IR fluxes
are taken from the NED on-line archive.
The FIR is defined as a
linear combination of 60 and 100$\mu$m
luminosities after David et al. (1992).
No significant difference between the FIR
distribution functions is observed on the basis of
the nuclear dust content. In vain we searched also for correlations
between the nuclear dust morphology and the host galaxy morphology
or the core 5~Ghz radio power (the plots of the corresponding
distribution functions are not shown). However, 
``dust-poor'' objects seem to correspond on average to intrinsically
more active objects, the average of the [O{\sc iii}] luminosity being
three times higher than in ``dust-rich'' objects [$1.2 \pm^{0.5}_{0.3}$
versus $(0.42 \pm^{0.13}_{0.10}) \times 10^{42}$~erg~s$^{-1}$].
Again, future enlargements of the sample, for
which a classification of the nuclear dust morphology is available, would be
of the uppermost importance to confirm this hint.

\section{The anticorrelation between Balmer Decrement and X-ray absorption}

Our analysis shows that the X-ray absorbing column density
and the Balmer Decrement are anti-correlated in Seyfert 2S along
the whole range of ${\rm N_H}$
between $10^{21}$ and $10^{24}$~cm$^{-2}$.
Fig.~\ref{fig2} and \ref{fig7}
suggest that the higher the X-ray column density,
the lower
the amount of reddening of the optical lines.
The latter is a sensitive measure of the amount of dust-driven extinction.
The intrinsic spectrum of hydrogen recombination lines is well known, for the
conditions typical of the low-density Narrow Line Regions (NLR), and
the derived optical extinction is not strongly dependent on the exact
extinction law adopted.
That dust is a common ingredient of NLR has been suggested by several
studies (Mac Alpine 1985; Osterbrock 1989). Netzer \& Laor
(1993) suggested that the apparent lower covering fraction of the
NLR than the
Broad Line Regions (BLR), the ``emission gap'' between NLR and BLR and the
scaling of the BLR size with luminosity (${\rm R_{BLR} \propto L^{1/2}}$)
may be all explained if dust is embedded in the line emitting gas,
and the BLR is located within the sublimation radius. The correlation
in Fig.~\ref{fig2} and Fig.~\ref{fig7} is therefore likely to be telling us
something about the physical structure and spatial distribution of the
matter responsible for the non-stellar ionization-driven emission lines.

The observed anticorrelation between Balmer decrement and ${\rm N_H}$ implies
that the higher the amount of X-ray absorption, the more dust-free the
absorbing matter is.
We suggest
two possible interpretations.
First, let's suppose that the matter responsible for the
X-ray and optical obscuration is the same.
Given a constant amount of mass available for nuclear absorption, distributed
in an spherical shell, the observed ${\rm N_H}$ scales as the inverse
square of the inner side of the shell. Otherwise stated,
the higher ${\rm N_H}$,
the longer the integration
path, the closer to the center one goes, and consequently the brighter
the ionizing continuum is seen by the obscuring matter,
thanks both to the lower distance from the
nuclear source and to the lower absorption of the interposed
obscuring layers. The absorbing matter could hence extend towards the center
up to radius where the dust is significantly sublimated and therefore 
not contribute substantially to the optical reddening of the narrow lines.
The sublimation radius is ${\rm \sim 0.02 L_{45}^{1/2}}$~pc, where ${\rm
L_{45}}$ is the nuclear luminosity is units of $10^{45}$~erg~s$^{-1}$
(Laor \& Drain 1993). This is very close to the magnitude and
luminosity scaling of the BLR radius (Netzer 1990; Clavel et al. 1991;
Peterson et al. 1991). Such an evidence
suggests that the BLR is actually mostly dust-free.
This scenario, however, leaves unexplained why the properties of the
absorbing matter on sub-parsec scales are connected with the NLR
on two order of magnitude larger scales (Axon et al. 1997).

Alternatively, the optical and X-ray absorbing media may be totally
decoupled, and the
X-ray absorber may almost dust-free and/or located
more innermost than the NLR (the latter hypothesis is in agreement
with the unification scenario, and observationally supported
in the Circinus~Galaxy; Oliva et al. 1994; Matt et al. 1999).
The optical reddening might be due to
more external matter, associated for example either with the galactic
disk or with the dusty
structures seen in M98 images. The observed correlation might
imply an ``evolutionary correlation'' between these two media, in the
sense that higher-${\rm N_H}$ X-ray absorbing matter could
by yielded through depletion of
the nuclear environment dust, therefore yielding {\it indirectly} a lower
reddening of the optical lines. It is admittedly not straightforward
to envisage a physical process, responsible for the ``transfer''
of matter from hundreds to a few pc scale. Dynamical perturbations as
traced by stellar bars may play a r\^ole. This point is further
discussed in Sect.~6.

Whatever the origin of such a correlation is, it is no more
valid in Compton-thick objects.
In most of these objects, only lower limits on the X-ray absorbing
column density can be measured. However, the corresponding Balmer
decrement values
span basically the same
range as the in Compton-thin objects, while they should
cluster at very low values if the correlation were
true for them as well.
Again, this suggests that the optical and X-ray absorbers are
totally decoupled. Moreover,
this represents,a s far as we know, the first
observational evidence that the
X-ray absorbing matter in Compton-thin and -thick objects is qualitatively
different.

\section{On the dust morphology classes}

A possible correlation of the nuclear dust morphology classification
of M98 with the amount of X-ray absorption,  suggests
that the ``Compton-thinness'' is related
to the nuclear dust morphology on the hundreds of parsecs scale.
The Compton-thin
objects exhibit preferentially ``dust-rich'' nuclear environment, while the
Compton-thick can indifferently have dust-rich or dust-poor environments.
In Compton-thick Seyfert~2s the X-ray absorption might therefore be totally
decoupled from the dust content of the nuclear environment.
The presence of nuclear dust is not correlated with the galaxy type
or with the FIR luminosity. It corresponds to a slightly
(about a factor of three) higher average [O{\sc iii}] luminosity.
It would be very interesting to search for a confirmation
of this X-ray absorption/morphology connection. The high-throughput
scientific payload on board XMM-Newton may soon
provide invaluable further measurements. Despite the still
marginal evidence, we don't refrain from speculating a bit on this
intriguing hypothesis.

One may suppose that Compton-thick objects are those, which indeed
contain the dusty homogeneous molecular ``torus'', envisaged by the
Seyfert unification theories (Antonucci \& Miller 1985). The innermost
side of this torus would
be located at distances of the order of 1~pc (or fractions,
as suggested by the water maser measurements in NGC~1068, Greenhill et
al. 1997). The formation history of this ``standard'' torus cannot be
inferred from our results alone, but, irrespectively of the details, it could
contribute to deplete the dust present in the nuclear
environment on $\sim$100~pc scale. Recently, Maiolino et al. (1999) pointed
out that Seyfert~2s with stellar bars tend to show the highest fraction of
Compton-thick objects. Bars may therefore be very efficient in driving
toward the nuclei the matter, which is responsible for the bulk of the
obscuration, as already suggested on theoretical basis by
Barnes \& Henquist (1995).
Being this true, the formation of a compact torus might be
related to the strength of the central gravitational potential, as suggested
by the average higher [O{\sc iii}] luminosity of ``dust-poor'' versus
``dust-rich'' objects.
However, there is no strong evidence so
far that bars are preferentially found in active nuclei (Heckman 1980;
Simkin et al. 1980; Mulchaey \& Regan 1997; M98) and recent
high-resolution imaging observations of a small sample
of Seyfert galaxies with HST rule out that nuclear bars are the primary
fueling mechanism for Seyfert nuclei (Regan \& Mulchaey 1999).

On the other hand, Compton-thin sources
would not have the compact torus in their innermost regions and might
therefore keep their nuclear environment enough ``dust-rich'' for this
dust to be detected in the high-resolution images of M98. We stress that
there is no evidence that the dust lanes seen in HST images are {\it themselves}
responsible for the X-ray absorption. Significant column densities are measured
not only in sources where the nucleus is covered by dust (which are actually
a few), but also
in objects where the nuclear environment show dust lanes and patches, which
do not intercept our line of sight to the nucleus. The above correlation
is therefore valid in ``environmental terms'', suggesting, otherwise stated,
that the X-ray absorption measured on the Earth is the superposition of several
individual unresolved
``clouds'', which are statistically more numerous in Seyfert~2s than in
Seyfert~1.

The above results are consistent with a scenario (as the GDM),
where the X-ray obscuration, that turns a Seyfert~1 into a Seyfert~2,
occurs in the host galaxy on scales of the order of hundreds parsecs.
This scenario has been suggested by several authors in the past (Maiolino \&
Rieke 1995; Mcleod \& Rieke 1995; Simcoe et al. 1997; M98). Recently,
Matt (2000), has elaborated an extension of the Seyfert
unification models, where Compton-thick Seyfert 2s are observed
through compact, thick matter with a large covering factor and
close (a few tens parsecs at most) to the nucleus, while Compton-thin
or intermediate Seyferts are obscured by dust lanes at larger distances.
Whether these two classes, unveiled to be qualitatively
different and not simply the manifestation of different values of a
critical observable (${\rm N_H}$), are evolutionary
linked in a sequential scenario, or they are created ``in parallel''
from the same parents,
according to a hidden critical parameter (the central black hole mass?) is
an open question for the models of active galaxy evolution.

\section{Conclusions}

In this {\it paper}, we studied the correlation between X-ray and optical
absorption, the latter measured through the Balmer decrement of the
narrow hydrogen recombination lines. Moreover, we extended this
study by encompassing the nuclear dust morphological classification by
M98. The main results of our study can be summarized as follows:

\begin{itemize}

\item[a)] in Compton-thin Seyfert~2s ({\it i.e.}: those with ${\rm N_H \simlt
10^{24}}$~cm$^{-2}$), the Balmer decrement and the amount of X-ray absorption
are strongly {\it anti-}correlated along three orders of magnitude
in ${\rm N_H}$, the residual scattering in the correlation being easily
explained with still pending uncertainties in the measure of ${\rm
N_H}$ and/or in long-term variations of the
nuclear flux in a few individual objects \\

\item[b)] this correlation does not hold for Compton-thick objects
(${\rm N_H \simgt 10^{24}}$~cm$^{-2}$), which span a comparable range in
Balmer decrement as the Compton-thin, despite the much higher amount
of X-ray absorption \\

\item[c)] Compton-thin Seyfert 2s seems to prefer nuclear environments,
which are
rich of dust on the hundreds parsecs scale. On the other hand, Compton-thick
Seyferts do not share the same behavior, exhibiting indifferently
``dust-poor'' and ``dust-rich'' environments. The statistical significance of
this results is still marginal, but can be easily improved in the
nearby future with an
enlarging of the sample of Seyfert~2 galaxies for which {\it both} high
resolution images of the nuclear regions {\it and} reliable measures of
the X-ray absorbing column density are available \\

\end{itemize}

\section*{Acknowledgments}

MG acknowledges an ESA Research Fellowship. This research has made use
of the NASA/IPAC Extragalactic Database, which is operated by the Jet
Propulsion laboratory under contract with NASA. We gratefully acknowledge
R.Maiolino for enlightening discussions and comments.

{}

\end{document}